\documentstyle[12pt]{article}

\def\ra{\rightarrow}
\def\be{\begin{equation}}
\def\ee{\end{equation}}
\def\bea{\begin{eqnarray}}
\def\eea{\end{eqnarray}}
\def\g{\gamma}
\def\G{\Gamma}
\def\s{\sigma}
\def\S{\Sigma}
\newcommand{\lb}{(l^+l^-)_b}
\newcommand{\e}{e^+e^-}
\newcommand{\EE}{e^++e^-}
\newcommand{\eb}{(e^+e^-)_b}
\newcommand{\m}{\mu^+\mu^-}
\newcommand{\MM}{\mu^++\mu^-}
\newcommand{\mb}{(\mu^+\mu^-)_b}
\newcommand{\meb}{(\mu^+e^-)_b}
\newcommand{\emb}{(\mu^-e^+)_b}
\newcommand{\ta}{\tau^+\tau^-}
\newcommand{\TT}{\tau^++\tau^-}
\newcommand{\tb}{(\tau^+\tau^-)_b}
\newcommand{\n}{n^3S_1}
\newcommand{\p}{\pi^0\gamma}

\begin{document}

\title{ON QUASIPOSITRONIA}

\author{A. A. MALIK and I. S. SATSUNKEVICH
\\{\em B.I. Stepanov Institute of physics}
\\{\em National Academy of Sciences of Belarus}
\\{\em F.Skoryna avenue 68, Minsk 220072, Belarus}}

\date{}

\maketitle
\begin{abstract}
Quasipositronia are atoms of simple hydrogenlike structure. They are
not included in periodic table of elements. Examples of such atoms
are positronium $\eb$, muonium $\meb$, antimuonium $\emb$, $\mu$ and
$\pi$ mesoatoms, $(\pi^+\pi^-)_b$, $(\pi K)_b$, $\mb$ and $\tb$.
Such atoms as $\eb$, $\mb$, $\tb$, $\meb$ and $\emb$ are specially
interesting from the point of view of investigations of fundamental
properties of matter. Their description is rather simple and in many
aspects similar to each other.
In this paper energy levels and decay widths of different decay
channels of $\mb$ and $\tb$ has been given. Cross section of
production of these atomic systems in $\e$ annihilation has been
calculated. To search these atoms, required experimental conditions
for initial beams and detectors has been found.
\end{abstract}

\section{Introduction}

QED is a well known theory to describe electromagnetically bound
atomic states. A more general theory than QED which gives an overall
framework for the description of all quasipositronia, was
constructed in \cite{CL} by integrating out from QED all the degrees
of freedom with energies of the order or above the mass of the
constituents.

We use this theory in the first approximation on $\alpha$
($\alpha\simeq 1/137$ is the constant of fine structure) and
$m^{-1}$. As in the case of hydrogen atom the hamilton operator for
$\lb$ atoms is \be H={{\bf P}^2\over m_l}-{\alpha\over r}, \ee where
${\bf P}=-i\partial/\partial{\bf r}$, $r=|{\bf r}|$ is distance
between lepton and antilepton. This operator differs from the
hydrogen atom hamiltonian only by reduced mass. In hydrogen atom
hamiltonian, reduced mass is just the mass of electron but in case
of $\lb$ atom reduced mass is $m_l/2$ where
$m_l=m_e,m_{\mu},m_{\tau}$, so the energy levels of such systems are
obtained easily using the corresponding reduced masses. Different
corrections to the hamiltonian {\em eg.} relativistic mass growth,
orbital, spin-orbital, spin-spin and annihilation interactions, all
give the hyperfine structure levels. Perturbation hamiltonian
contains only the sum of spins of particles, so the energy levels
are divided into singlets (parastates with total spin 0) and
triplets (orthostates with total spin 1).

$\mb$ and $\tb$ are interesting experimentally because they have
never been produced in laboratory. Their production and study is
also interesting from the point of view of QED tests, in
comparison with the hydrogen atom they don't have hadron nucleus
with all it's theoretically indefinite behavior. They differ from
$\eb$ and $\meb$ due to the new decay channels.

\section{Energy levels and decay channels of quasipositronia}

$\eb$ and $\meb$ are well known bound systems. $\mb$ and $\tb$ are
bound states similar to $\eb$ but they are difficult to produce. A
special $\e$ collider at energy 211 MeV is needed to produce $\mb$
in huge quantities. The atomic system which consists of coulombic
bound state of a $\tau^{-}$ and $\tau^{+}$ can also be produced in
$\e$ annihilation just below $\tau$ pair threshold \be
\EE\ra\g_{virtual}\ra\lb\label{eq:eellb} \ee The $\tau$-charm
factories can offer the best route for making these atoms, but there
is also a rather appealing way to make them just now at the Beijing
Electron Positron Collider (BEPC) using Beijing Spectrometer
(BESII), and as we guess in near future at VEPP-4M in Novosibirsk.

The energy levels of the $\lb$ atom are given as for hydrogen atom
with a reduced mass ${m_l/2}$ \be E_n=-{m_l\alpha^2\over 4n^2}. \ee
Knowing the energy spectra of these atoms now we need to consider
different decay channels of $\lb$ systems.

There are two classes of decay channels \cite{Per}. In the first
class the $l^-$ or $l^+$ ($l=\mu,\tau$) decay through the weak
interaction in the normal way and the atomic state disappears. The
decay width is \be \G(\lb,l\ decay)={2\over\tau_l} \ee where
$\tau_l$ is lifetime of lepton. Numerical values of all decay widths
are given in table 1.

In the second class of decay channels the $l^-$ and $l^+$ annihilate.
The annihilation requires that the atomic wave function $\Psi (r)$ be unequal
to 0 at r=0 {\em i.e.} $\Psi(0)\neq0$. Here r is the distance between the
$l^-$ and $l^+$. Therefore in lowest order, annihilation occurs only
in L=0 state, {\em i.e.} S state. The annihilation channels of $n^{3}S_{1}$
state of $\lb$ with corresponding decay widths are following:
The channel
\be
\lb\ra\g+\g+\g
\ee
has the width \cite{Ber}
\be
\G(\lb\ra3\g)={2(\pi^2-9)\alpha^6m_l\over9\pi n^3}
\ee
The two channels
\be
\lb\ra\EE
\ee
\be
\tb\ra\MM
\ee
have the same width
\be
\G(\lb\ra\EE)={\alpha^5m_l\over6n^3}
\ee
Finally there is hadron channel
\be
\lb\ra hadrons
\ee

For $\tb$ we can calculate the width of this channel using colliding beams
$\e$ annihilation data at $E_{tot}\sim2m_{\tau}$
\be
\s(\EE\ra hadrons)\approx2\s(\EE\ra\MM)
\ee
Therefore
\be
\G(\tb\ra hadrons)\approx2\G_{\tau ee}
\ee
Neglecting $\G(\tb\ra3\g)$ we get the total width
\be
\G_{\tau}\approx\G(\tb,\tau\ decay)+4\G_{\tau ee}=\Biggl(4.5+{24.5\over n^{3}}\Biggr)\times10^{-3}eV
\ee

In case of $\mb$ there is only one hadron channel \cite{MSV} \be
\mb\ra\g+\pi^0 \ee The width can be calculated in two ways. The
first one is phenomenological. As we know \cite{Aug} the quark
production cross section in $\e$ annihilation is very small in the
range of $\mb$ threshold energy. \be \s(\e\ra u\bar u\mbox{ or
}d\bar d)\sim\s(\e\ra \pi^+\pi^-)\simeq40nb \ee Production of a
photon in $\e\ra(u\bar u,d\bar d)+\g$ changes the result by a
factor $\alpha$, so we have \be \s(\e\ra\p)\sim0.3nb. \ee If we
compare this value with the cross section of $\e\ra\m$, which
equals to 2.2 $\mu$b above threshold energy, we see that we can
neglect $\s(\e\ra\p)$. So phenomenological result is very small:
\be \G(\mb\ra\p)\sim{10^{-7}\over n^3}eV \ee In the second method
we use the cross section of $\e\ra\p$ from \cite{Bra} \be
\G(\mb\ra\p)=4\s(\sqrt s = 211
MeV)\S|\psi_n(0)|^2\simeq3.4\times10^{-9}eV \ee This decay width
is very small. So the total width $\G_{\mu}$ is defined by the
$\e$ channel mainly.

\begin{tabular}{|l|c|c|}
\multicolumn{3}{c}{Table 1: Decay widths of $\mb$ and $\tb$ atoms}\\
\hline
&$\mb$&$\tb$\\
\hline
$\tau_l$&$2.20\times10^{-6}s$&$2.90\times10^{-13}s$\\
$\G(\lb,l\ decay)$&$5.99\times10^{-10}eV$&$4.54\times10^{-3}eV$\\
$\G(\lb\ra3\g)$&$1.18\times10^{-6}eV$&$1.99\times10^{-5}eV$\\
$\G(\lb\ra\e)$&$4.39\times10^{-4}eV$&$7.37\times10^{-3}eV$\\
$\G(\lb\ra\m)$&--&$7.37\times10^{-3}eV$\\
$\G(\lb\ra hadrons)$&$3.4\times10^{-9}eV$&$1.48\times10^{-2}eV$\\
Total decay width&$4.4\times10^{-4}eV$&$3.4\times10^{-2}eV$\\
\hline
\end{tabular}
\\Using these values we can calculate production cross section of
$\lb$.

\section{Production of the $\lb$}

The production cross section of the process (\ref{eq:eellb}), according to the
Breit-Wigner equation, is
\be
\s(\EE\ra\lb)={3\pi\over4m_l^2}{\G_{lee}\G_l\over(E-2m_l)^2+\G_l^2/4}
\ee
Here E is the total energy of $e^{-}$ and $e^{+}$. The peak cross section is
\bea
\s(\EE\ra\lb,peak)&=&\sum_n{3\pi\over m_l^2}{\G_{lee}\over\G_l},\\
\s(\EE\ra\mb,peak)&=&0.33b,\\
\s(\EE\ra\tb,peak)&=&0.25mb.
\eea
Taking into account the radiative corrections \cite{Bay} we get
\be
\s_r=\s\exp\Biggl(-{4\alpha\over\pi}ln{m_l\over m_e}ln{m_l\over\G_l}\Biggr).
\ee

Now we consider the possibility that initial $e^{-}$ or $e^{+}$ can
radiate soft photon with \be {\omega\over m_l}\ll1. \ee The criteria
of perturbative theory application in this case are \be \alpha
ln{m_l\over m_e}\ll1,\hspace{1cm} {\alpha\over\pi}ln{m_l\over
m_e}ln{m_l\over\omega_{min}}\ll1.\label{eq:cond} \ee From
(\ref{eq:cond}) it implies that we can put $\omega_{min}$ equal to
the decay width of $1^{3}S_{1}$ state of $\lb$. The cross section of
radiative production of $\lb$ is \be
\s_{rad}=\sum_n{\pi\over2n^3}{\alpha^6\over
m_l\omega_{min}}\Biggl(2ln{m_l\over m_e}-1\Biggr). \ee

As the electrons and positrons in accelerator beams have some energy spread so
we should use an average value of cross section \cite{Cab}
\be
<\s>=\int\s(E)\rho(E)dE.
\ee


$\mb$ is produced in the $\e$ annihilation through $\n$ states,
which decay into $3\g,\e$ or $\p$. Peak cross section of this process is
0.33 b. This cross section is very big if the decay width is small. In
this case we should take into account the factors which decrease
the effective cross section. At first we should consider radiative
corrections due to soft photon radiation and then average cross section
over the energy spread of $e^-$ and $e^+$
beams. Usually the energy spread $\triangle$E, is much greater
than $\G$. All the working accelerators have energy spread of a
few MeV. For such an accelerator it is quite difficult to observe
production and decay of $\mb$. But for a special $\e$ collider
with $\sqrt s$=211MeV and beam's energy spread less than 1keV, we
have \cite{Bay}
\be
<\s_r>=0.16\pi^2{\alpha^5\over m_{\mu}\triangle E}\label{eq:sr},
\ee
\be
<\s_{rad}>=3.6{m_{\mu}\over\triangle E}\Biggl(ln{\triangle E\over
\omega_{min}}-1\Biggr)\times10^{-14}b,
\ee
$\s_r$ is Breit-Wigner cross section of $\mb$ production with the account
of virtual radiative corrections, $\s_{rad}$ is cross section of
radiative production of $\mb$ with the emission of a real photon
having energy up to $\omega_{min}\sim\G_{\mu1}$ = 3.65$\times10^{-4}$
eV. The total differential cross section of the production and
decay of $\mb$ is
\be
{d\s\over d\Omega}={<\s_r>+<\s_{rad}>\over4\pi}\simeq16nb/ster\label{eq:ds}
\ee
where $\Omega$ is solid angle. Comparing this number with the cross section
1.1$\mu b/ster$ of elastic scattering at $\theta=90^{\circ}$
we find that the birth of atomic $\mb$ increases the elastic cross section at
$90^{\circ}$ by 1.5$\%$, which is a detectable quantity in an exact experiment.


Now we consider the production of $\tb$ at BEPC. Let the electrons
and positrons are equally distributed in some energy interval
$\triangle E$ near threshold. At BEPC we have $\triangle E=1.47MeV$
\be <\s_r>=\s_r{\pi\over2}{\G_{\tau ee}\over\triangle E}=0.56pb \ee
\be <\s_{rad}>=\s_{rad}{\omega_{min}\over\triangle
E}\Biggl(ln{\triangle E\over\omega_{min}}-1\Biggr)=10.89pb \ee
Number of events which can be observed at BEPC having luminosity
$L=0.361pb^{-1}$ are \cite{MSQS} \be
N=(<\s_r>+<\s_{rad}>)L=11.45\times0.361=4.13events \ee The cross
section $\s(\e\ra\ta)$ just below $\tau$ pair threshold at 3.5538
GeV, with the account of all the radiative corrections is, according
to Kirkby \cite{Kir} \be \s(\EE\ra\TT)=110pb \ee Branching ratio for
$\tau\ra\mu\nu\nu$ is 0.1735, so we have \be
\s(\EE\ra\TT\ra\MM)=110\times(0.1735)^2=3.3pb \ee Branching ratio
for $\tb\ra\m$ is 0.21, so the cross section is \cite{MSG3} \be
\s(\EE\ra\tb\ra\MM)=11.45\times0.21=2.4pb \ee Such a big cross
section is due to the branching ratio for $\tb\ra\MM$ which is 0.21
in spite of 0.03 as for free $\ta$ pair. The ratio between the cross
section of bound state $\tb$ and free $\ta$ pairs is \be
R={2.4\over3.3}=0.73 \ee According to BEPC \cite{Bai} at 3.5538 GeV
for the luminosity of 0.361 $pb^{-1}$ there is 1 $e\mu$ event but
there should be 1.73$\m$ events and 1.73 $\e$ events. The increment
0.73 is due to $\tb$.

\section{Conclusion}

It follows from our estimates for $\tb$ decay rates and production
cross section that $\tb$ decays give a rather big number for $\m$
and $\e$ events at BEPC. It is a pity that at BEPC the conditions
for events registration are such that $\m$ and $\e$ decay channels
of $\tb$ is impossible to register. Another possibility to produce
$\tb$ is VEPP-4M in Novosibirsk.

$\mb$ we can produce only in a specialized $\e$ collider with $\sqrt
s$=211MeV and beam energy spread less than 1keV.

\end{document}